\documentclass[ ]{aa}
\usepackage{graphicx}
\usepackage{color}
\usepackage{epstopdf}
\usepackage[round]{natbib}
\usepackage{graphics}
\usepackage[figuresright]{rotating}
\usepackage{amssymb}
\usepackage{bm}
\usepackage{epsfig}
\usepackage{color}
\usepackage{amsmath}

\begin{document}

%%%%%%%%%%%%%%%%%%%%%%%%%%%%%%%%%%%%%%%%%%%%%%%%%%%%%%%%%%%%%%%%%%%%%%
%%% Begin of Document
%%%%%%%%%%%%%%%%%%%%%%%%%%%%%%%%%%%%%%%%%%%%%%%%%%%%%%%%%%%%%%%%%%%%%%

\title{Testing Dark Energy Models through Large Scale Structure}

\author{Olga Avsajanishvili\inst{1}, Lado Samushia\inst{2,1}, Natalia A. Arkhipova\inst{3},  and Tina Kahniashvili\inst{4,5,1}
}
\institute{Abastumani Astrophysical Observatory, Ilia State University, 3-5 Cholokashvili Ave., Tbilisi, 0194, Georgia
\and
Department of Physics, Kansas State University, 116 Cardwell Hall, Manhattan, KS, 66506, USA\\
\and
Astro Space Center of P.~N.~Lebedev Physical Institute, Russia, 84/32 Profsoyuznaya str.,
Moscow, 117997 Russia\\
\and
McWilliams Center for Cosmology and Department of Physics, Carnegie Mellon University, 5000 Forbes Ave, Pittsburgh, PA 15213 USA\\
\and
Department of Physics, Laurentian, University, Ramsey Lake Road, Sudbury, ON P3E 2C, Canada\\
}

%\email{olga.avsajaniashvili@iliauni.edu.ge}
%email{lado@phys.ksu.edu}
%\email{arna@asc.rssi.ru}
%\email{tinatin@andrew.cmu.edu}

%%%%%%%%%%%%%%%%%%%%%%%%%%%%%%%%%%%%%%%%%%%%%%%%%%%%%%%%%%%%%%%%%%%%%%%%%%%%%%%%%%%%%%%%%%%%%%%%%%%%%%%%%%

\date{\today}

%%%%%%%%%%%%%%%%%%%%%
%%% Content
%%%%%%%%%%%%%%%%%%%%%

%%% Abstract

\abstract{This paper is a contribution to the Proceedings
of the 5th Gamow International Conference, Odessa, Ukraine, August 2015.
%This paper is based on the \citep{av14}.

We explore the scalar field quintessence freezing  model of dark energy with the
inverse Ratra-Peebles potential. We study the cosmic expansion and the large
scale structure growth rate. We use recent measurements of the growth rate and
the baryon acoustic oscillation peak positions to constrain the matter density
$\Omega_\mathrm{m}$  parameter and the model parameter $\alpha$ that describes
the steepness of the scalar field potential.} {To study  the background dynamics
of the the $\phi$CDM model. To investigate the influence of the scalar field on
the expansion rate of the Universe.  To examine the background evolution of the
$\phi$CDM model on the cosmological model parameters and on the content of the
Universe.   To study  the equation of state parameter ${\rm w}(a)$ (with the
scale factor $a$) during the expansion of the Universe.  To investigate the
influence of the $\phi$CDM model on evolution of the large scale structure. To
explore the applicability of the Linder  $\gamma$-parametrization of  growth
rate for the $\phi$CDM model and to define a redshift range of this
parametrization validity. To derive observational constraints on  the model
parameters $\Omega_\mathrm{m}$ and $\alpha$.} {We solve jointly the equations
for the background expansion and for the growth rate of matter perturbations.
The obtained theoretical results are compared with the observational data.  We
perform the Baysian data analysis to derive constraints on the model
parameters.} {The larger value of the $\alpha$ parameter implies stronger time
dependence of the scalar field. For the Ratra-Peebles $\phi$CDM model the
expansion of the Universe occurs faster with increasing of the $\alpha$
parameter.  The scalar field begins to be dominant earlier with the increasing
value of the $\alpha$ parameter. The Ratra-Peebles $\phi$CDM model predicts a
slower growth rate than the $\Lambda$CDM model. The Linder
$\gamma$-parametrization works well for the Ratra-Peebles $\phi$CDM model in the
range of the redshifts  ${\rm z}\in(0;5)$.}{}


\titlerunning{Testing Dark Energy Models through Large Scale Structure}

\authorrunning{Avsajanishvili et al.}

\maketitle
%\footnote {olga.avsajaniashvili@iliauni.edu.ge}
%%%%%%%%%%%%%%%%%%%%%%%%%%%%%%%%%%%%%%%%%%%%%%%%%%%%%%%%%%%%%%%%%%%%%%%%%%%%%%%%%%%%%%%%%%%%%%%%%%%%%%%%%%
\section{Introduction}

According to the cosmological observations our Universe expands with an
acceleration \citep{perlmutter99,riess98,riess07}.  There are several models for
explanation of this phenomenon. The most common approach is to assume that order
of $70\%$ of the energy density of the Universe is present in the form of dark
energy (DE).

The simplest model of DE admits that DE is the vacuum energy, that is given in
the form of the time-independent cosmological constant $\Lambda$. This model is
referred to as a {\it concordance} model since it is in a very good agreement
with all available today cosmological observations.  The $\Lambda$CDM model,
however, suffers from the fine tuning and the coincidence problems
\citep{carroll00,pad02,pr03,martin2012}. To alleviate these problems, other
models of DE have been proposed
\citep{cds97,adm99,amendola00,wetterich95,kmp01,cct,dgp,shi15,duniya15,chen15,pradhan13}.  The main alternatives
of the $\Lambda$CDM model are the models involving a dynamical scalar field, so
called the $\phi$CDM models. This family of models avoids some theoretical
foundation difficulties, namely: the fine tuning problem, having a more natural
explanation for the observed low energy scale of DE \citep{zws99}.  For the
$\Lambda$CDM model the equation of state parameter ${\rm w}$ is a constant and
is equal to minus one. For the $\phi$CDM models the equation of state  ${\rm w}$
parameter is time dependent (we will use below the scale factor factor function
${\rm w}(a)$) and approaches to minus one today, i.e. ${\rm w}(a_0) \rightarrow
-1$, here $a_0$ is the today value of the scale factor normalized to be one,
$a_0 \equiv a_{\rm today}=1$ \citep{cl05,yw12}.

Depending on the value of the equation of state ${\rm w}$ parameter today, the
$\phi$CDM models are divided into two classes: the phantom models
($-{1}/{3}<{\rm w}<{-1}$) and the quintessence models (${\rm w}>{-1}$). The
quintessence models are subdivided into the thawing models, for which the
evolution of the scalar field is fast \citep{scherrer07,linder2015,lima15} and the
tracking (freezing) models, for which the evolution of the scalar field is slow,
compared to the Hubble expansion \citep{cds98,cl05,lk09,dl08,cdt13}.  In the tracking
models, the scalar field has a tracking solution, in which the scalar field
energy density, remaining subdominant, tracks at first the radiation and then
the matter energy densities \citep{bm02,swz99}. At late times, the scalar field
becomes the dominant component and starts to behave as a component with the
effective negative pressure,  that leads at the late stages to the accelerated
expansion of the Universe.  The simplest representation of such a model is the
scalar field model, when the scalar field potential is given through an inverse
power law, Ratra-Peebles, potential \citep{rp88}.  We will refer to this model
below as the $\phi$CDM Ratra-Peebles model.

The background expansion history (as well as the large scale structure growth
rate) is different for the scalar field ($\phi$CDM) model and for the
concordance ($\Lambda$CDM) model.  Thus the $\phi$CDM model can be distinguished
from the $\Lambda$CDM model through high precision measurements of distances and
growth rates over a wide redshift  range
\citep{spr12,pab12,psr12,bgs11,fwy10,ln10,pfmb11,akl09,dl08,st09}.

We present solutions of the joint equations for the background expansion and for
the growth rate of matter perturbations. We use a compilation of the recent
measurements of the growth rate and the baryon acoustic oscillation (BAO) to put
constraints on the $\alpha$ and the $\Omega_\mathrm{m}$ parameters.

This paper is organized as follows.  In Sec.~\ref{sec:theory} we  investigate
the dynamics and the energy of the Ratra-Peebles $\phi$CDM models.  In
Sec.~\ref{sec:parametr} we explore the parametrization of the equation of state
parameter ${\rm w}(a)$ in the Ratra-Peebles $\phi$CDM model by the different
models.  In Sec.~\ref{sec:growth} we study the influence of the  Ratra-Peebles
$\phi$CDM models on the growth rate. In Sec.~\ref{sec:observ} we present the
comparison of the obtained theoretical results with the observational data. We
discuss our results and conclude in Sec.~\ref{sec:conclusion}. We use the
natural units with $c= {\hbar}=1$ throughout this paper.

\section{ Background dynamics in the Ratra-Peebles $\phi$CDM model}
\label{sec:theory}

\subsection{Background equations}
We study the $\phi$CDM model with the Ratra-Peebles potential given by \citep{rp88}:
\begin{equation}
V(\phi)=\frac{\kappa}{2}M_\mathrm{pl}^2\phi^{-\alpha},
\label{eq:Potential1}
\end{equation}
where $M_\mathrm{pl}$ is the Planck mass, $\kappa$ is a model parameter, and
$\alpha$ is a positive constant, which determines the steepness of the scalar
field potential. Current observational data suggest that $\alpha$ can not be
larger than $\alpha \leq 0.7$ \citep{samushia09,psr12,fr13}.  A larger value of
the $\alpha$ parameter implies a stronger time dependence  of the  scalar field
potential $V(\phi)$.  In the limit of $\alpha$=0, the Ratra-Peebles $\phi$CDM model
reduces to the $\Lambda$CDM model.

The equation of motion for the scalar field is \citep{rp88,sahni02}:
\begin{equation}
\ddot{\phi}+3H{\dot\phi}-\frac{1}{2}\kappa\alpha M_\mathrm{pl}^2\phi^{-(\alpha+1)}=0,
\label{eq:KleinGordon}
\end{equation}
where an over-dot represents the derivative with  respect to a physical time
$t$, $H(a)= {\dot a}/a$ is the Hubble parameter, the scale factor is
$a=1/(1+{\rm z})$, and ${\rm z}$ is a redshift.

The energy density and the pressure of the scalar field are \citep{rp88,sahni02}
\begin{eqnarray}
\rho_\phi & = &\frac{M_\mathrm{pl}^2} {64\pi} \Bigl(\dot{\phi}^2 + \kappa M_\mathrm{pl}^2\phi^{-\alpha} \Bigr),
\label{eq:Rho}  \\
P_\phi & = & \frac{M_{\mathrm{pl}}^2}{64\pi} \Bigl(\dot{\phi}^2 - \kappa M_\mathrm{pl}^2\phi^{-\alpha} \Bigr),
\label{eq:P}
 \end{eqnarray}
and, the corresponding equation of state ${\rm w}$ parameter is,
\begin{eqnarray}
{\rm w} = \frac{{\dot\phi}^2 - \kappa M_{pl}^2\phi^{-\alpha}}{{\dot\phi}^2 + \kappa M_\mathrm{pl}^2\phi^{-\alpha}}.
\label{eq:w}
 \end{eqnarray}

The scalar field energy density parameter is defined by,
\begin{equation}
\Omega_{\rm \phi}(a) = \frac{1}{12H_0^2}(\dot{\phi}^2+\kappa M_{\rm pl}^2\phi^{-\alpha}).
\label{eq:omegafi}
\end{equation}
and the first Friedmann
equation for the Ratra-Peebles $\phi$CDM model in spatially-flat Universe
is:
\begin{equation}
E^2(a) = \Omega_{\rm r0}a^{-4} + \Omega_{\rm m0} a^{-3}   + \frac{1}{12H_0^2}(\dot{\phi}^2+\kappa M_{\rm pl}^2\phi^{-\alpha}),
\label{eq:Friedmann}
\end{equation}
where $E(a)= H(a)/H_0$, with $H_0=100 h {\rm km/s/Mpc}$ is a value of Hubble
parameter today; $\Omega_{\rm r0}$, $\Omega_{\rm m0}$, and $\Omega_{\rm \phi0}$
the  dimensionless density parameters  for radiation,  matter and DE at present
time. During the late stages of the Universe's expansion (after radiation-matter
equality) we can neglect the radiation term in the Eq.~(\ref{eq:Friedmann}). In
what follows we will assume fiducial values $\Omega_{\rm m0}=0.315$, $\Omega_{\rm
\phi0}=0.685$,  $h=0.673$ consistent with \citep{ade13}.

\subsection{Initial conditions}

We integrate the set of equations Eq.~(\ref{eq:KleinGordon}), Eq.~(\ref{eq:omegafi}), and
Eq.~(\ref{eq:Friedmann}) numerically, starting from
$a_{\rm in}=5\cdot10^{-5}$ to the present time $a_0=1$.
We assume the
following initial conditions for the scalar field amplitude and its time derivative,
\begin{eqnarray}
{\phi}_{\rm in} &=&\left[\frac{1}{2}\alpha(\alpha+2)\right]^{1/2}a_{\rm in}^{\frac{4}{\alpha+2}},
 \label{eq:phi0}
 \\
{\phi}_{\rm in}^\prime &=&\Bigl(\frac{2\alpha}{\alpha+2}\Bigl)^{1/2}a_{\rm in}^{\frac{2-\alpha}{2+\alpha}},
 \label{eq:dphi0}
 \\
{\kappa} &=&\Bigl(\frac{\alpha+6}{\alpha+2}\Bigl)
\Bigl[\frac{1}{2}\alpha(\alpha+2)\Bigl]^{\alpha/2},
\label{eq:kappa2}
 \end{eqnarray}
\noindent
where a prime denotes the differentiation with respect to scale factor $a$.
To obtain these initial conditions we use the fact that the expansion of the
Universe during matter and radiation domination epochs has a power-law form and
use the ansatz
\begin{equation}
a(t)=a_\star \Bigl(\frac{t}{t_\star}\Bigl)^n,~~~~~~~\phi(t)= \phi_\star \Bigl(\frac{t}{t_\star}\Bigl)^p
\label{eq:Power}
\end{equation}
where  $a_\star \equiv a(t_\star)$ and $\phi_\star \equiv \phi(t_\star)$ are the
scale factor and the scalar field values at $t=t_\star$. The index $n$ depends
on the dominant component driving the expansion of the Universe and is $n=1/2$
in radiation dominated epoch and $n=2/3$ in matter dominated epoch. We solve the
set of equations  Eq.~(\ref{eq:KleinGordon}), Eq.~(\ref{eq:omegafi}), and
Eq.~(\ref{eq:Friedmann}), during the radiation (and/or the matter dominated)
epochs, to obtain the general expressions for $\kappa$, and the scalar field
amplitude $\phi$, and its time-derivative $\dot{\phi}$ (which depend only on the
$\alpha$ parameter and value of the index $n$; the details are given in Appendix
A of \citep{av14}):
\begin{equation}
\phi = n\alpha(\alpha+2)^{1/2}\Bigl(\frac{a}{a_{\star}}\Bigr)^{2/n(\alpha+2)},
 \label{eq:phi_gen}
\end{equation}
%\begin{equation}
% \dot{\phi} = 2\Bigl(\frac{n\alpha}{\alpha+2}\Bigr)^{1/2}\Bigl(\frac{a}{a_\star} \Bigr)^{2/n(\alpha+2)}t^{-1};
% \label{eq:dphi_gen}
 %\end{equation}
 \begin{equation}
 \kappa=\frac{4n}{M_{\rm pl}^2 t_\star^2}
\Bigl(\frac{6n+3n\alpha-\alpha}{\alpha+2}\Bigl)[n\alpha(\alpha+2)]^{\alpha/2},
 \label{eq:kappa}
 \end{equation}
We set $t_\star=M_{\rm pl}^{-1}$ and obtain Eq.~(\ref{eq:phi0}),
Eq.~(\ref{eq:dphi0}), and Eq.~(\ref{eq:kappa2}) assuming the initial conditions
are set in radiation dominated epoch ($n=1/2$).
%\begin{equation}
%\kappa  =  \Bigl(\frac{\alpha+6}{\alpha+2}\Bigl)
%\Bigl[\frac{1}{2}\alpha(\alpha+2)\Bigl]^{\alpha/2},
%\label{eq:kappa2}
%\end{equation}

\subsection{The dynamics and the energy  of the Ratra-Peebles $\phi$CDM model.}
\begin{figure}[t]
\resizebox{0.4\textwidth}{!}{%
 \includegraphics{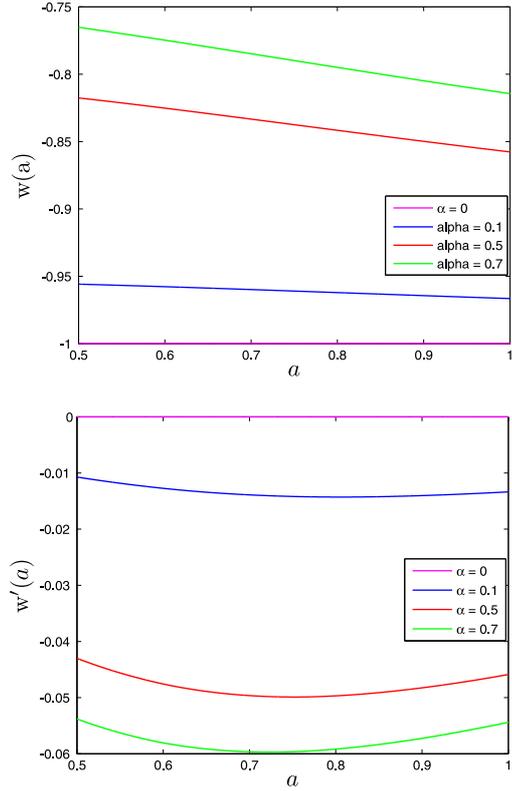}}
\caption{1a (upper panel) the DE equation of the equation of state parameter ${\rm w}(a)$ for the different values of $\alpha$ parameter and 1b (lower panel) the scale factor derivative ${\rm {w}^{\prime}}(a)$ of the equation of state parameter for the different values of the $\alpha$ parameter.}
\label{fig:f1}
\end{figure}

In this subsection we examine the evolution of the equation of state parameter
${\rm w}(a)$ and its scale factor derivative  ${\rm w}'(a)$  for  different
values of the $\alpha$  parameter (see Fig.\ref{fig:f1}).  The equation of state
parameter ${\rm w}(a)$  is a decreasing function of time (with the increasing
scale factor): that is a specific feature of the freezing models. In fact, a
large value of the $\alpha$ parameter results in a stronger time dependence for
the equation of state parameter ${\rm w}(a)$ (see Fig.\ref{fig:f1}a) and its
scale factor derivatives ${\rm w}'(a)$ (see Fig.\ref{fig:f1}b).

\begin{figure}[t]
\resizebox{0.4\textwidth}{!}{%
 \includegraphics{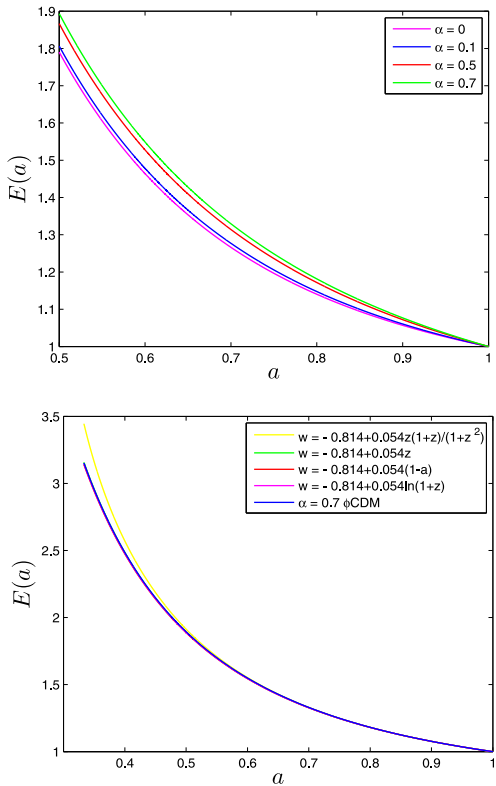}}
 \caption{2a (upper panel) the normalized Hubble expansion rate $E(a)$  for the Ratra-Peebles $\phi$ CDM model for  different values of the $\alpha$ parameter. 2b (lower panel) the normalized Hubble expansion rate $E(a)$ for BA, CH, GE, CPL parameterizations of ${\rm w}(a)$ and for the Ratra-Peebles $\phi CDM$ model, $\alpha=0.7$.}
 \label{fig:f2}
\end{figure}

Next, we investigate the influence of the scalar field $\phi$ on the expansion rate of the Universe. For the Ratra-Peebles $\phi$CDM model this expansion occurs faster with
increasing value of the $\alpha$  parameter  (see Fig.~\ref{fig:f2}a). The $\Lambda$CDM limit corresponds
to the slowest rate of the Universe's expansion.

\begin{figure}[t]
\resizebox{0.4\textwidth}{!}{%
 \includegraphics{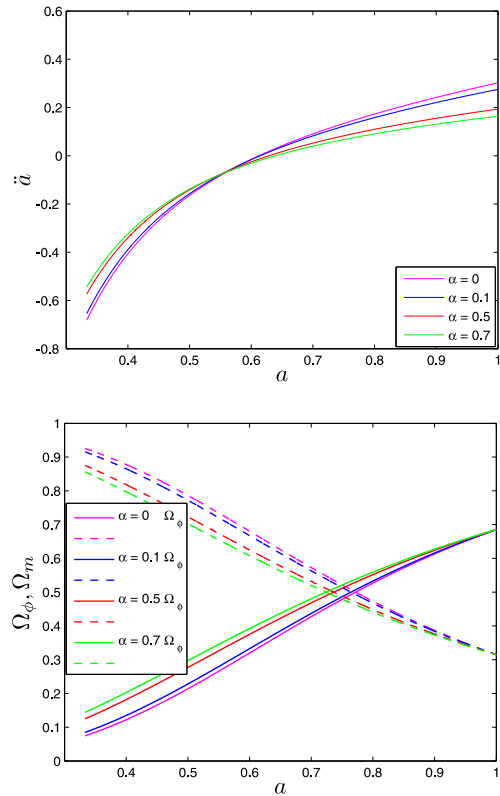}}
 \caption{ 3a (upper panel) the second derivative of the scale
factor  and 3b (lower panel) the matter energy density  $\Omega_{\rm m}(a)$ the  (dashed lines) and  the scalar field energy density $\Omega_{\rm \phi}(a)$ (solid lines) as functions
of the scale factor for different values of the $\alpha$ parameter.}
\label{fig:f3}
\end{figure}

We also study the background dynamics  (see Fig.~\ref{fig:f3}a), and the
evolution of the energy density of the matter and DE components, $\Omega_{\rm
m}$(a) and $\Omega_{\rm \phi}$(a) respectively  (see Fig.~\ref{fig:f3}b)  for the Ratra-Peebles $\phi$CDM model.
As we can see in the Ratra-Peebles $\phi$CDM model DE begins to be a dominant
component earlier than in the $\Lambda$CDM model, and the effect is stronger for
the larger value of the $\alpha$ parameter  (see Fig.~\ref{fig:f3}b), and thus the duration of the matter
dominated epoch becomes shorter.

\section{Parametrization of the equation of state parameter ${\rm w}(a)$ in the Ratra-Peebles $\phi$CDM model}
\label{sec:parametr}

There are several ways to parameterize the equation of state parameter ${\rm
w}(a)$, including:

\begin{itemize}
\item
 Coorkay and Huterer (CH) \citep{ch99}\\
  ${\rm w}(a) = {\rm w}_0+{\rm w}_a{\rm z},$
\end{itemize}

  \begin{itemize}
  \item
  Gerke and Estathiou (GE) parametrization \citep{efstathiou99}\\
  ${\rm w}(a) = {\rm w}_0+{\rm w}_a\ln(1+{\rm z}),$
 \end{itemize}

  \begin{itemize}
  \item  Chavalier and Polarsry, and Linder (CPL) parametrization \citep{cp01,linder03}\\ ${\rm w}(a) = {\rm w}_0+{\rm w}_a(1-a),$
  \end{itemize}

  \begin{itemize}
 \item
 Barboza and Alcaniz (BA) parametrization \citep{bja08} \\
  ${\rm w}(a)={\rm w}_0 + {\rm w}_a  \frac{{\rm z(1+z)}}{\rm{1+z^2}},$
\end{itemize}
where ${\rm  w}_0$ corresponds to the present day of the equation of state
parameter ${\rm w}(a)$, and ${\rm w}_a=({\rm d w}/{\rm d}{\rm z})|_{{\rm z}=0}=(-{\rm dw}/{\rm d}a)|_{a=1}$.

\begin{figure}[t]
\resizebox{0.4\textwidth}{!}{%
 \includegraphics{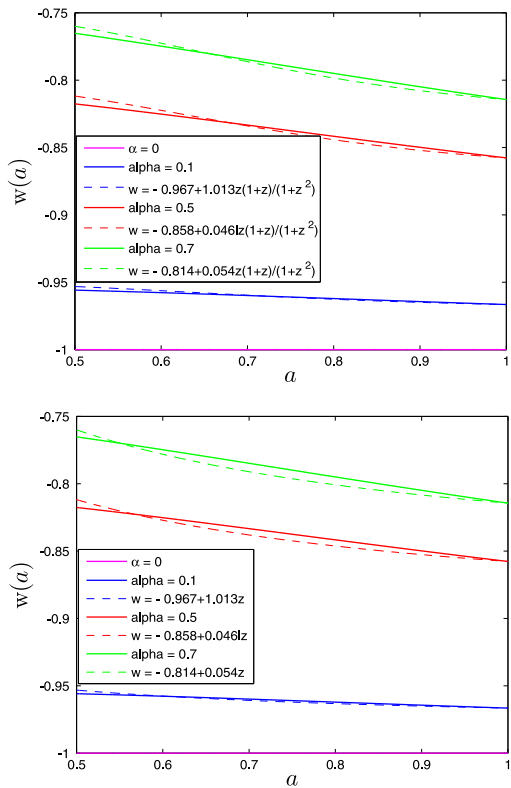}}
 \caption{4a (upper panel)  ${\rm w}(a)$ for different values of the $\alpha$ parameter along with predictions computed for the BA parametrization with corresponding
best-fit values for ${\rm w}_0$ and for ${\rm w}_{a}$. 4b (lower panel) ${\rm w}(a)$ for different values of the $\alpha$ parameter along with predictions computed for CH parametrization with corresponding best-fit values for ${\rm w}_0$ and for ${\rm w}_a$.}
 \label{fig:f4}
\end{figure}

\begin{figure}[t]
\resizebox{0.4\textwidth}{!}{%
 \includegraphics{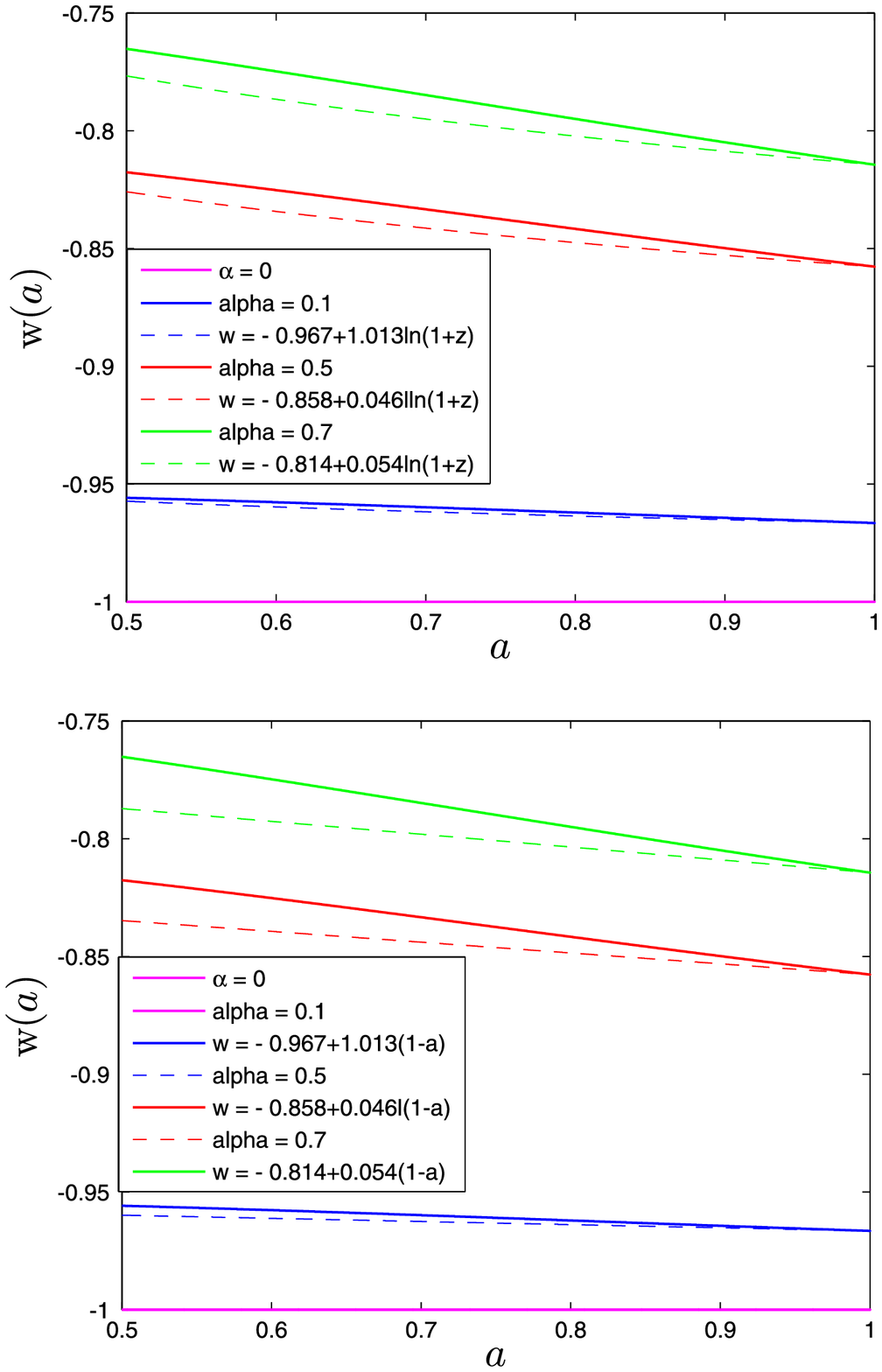}}
 \caption{5a (upper panel) ${\rm w}(a)$ for different values of the $\alpha$ parameter along with predictions computed for GE parametrization with corresponding
best-fit values for ${\rm w}_0$ and for ${\rm w}(a)$. 5b (lower panel) ${\rm w}(a)$ for different values of the $\alpha$ parameter along with predictions computed for the CPL parametrization with corresponding best-fit values for ${\rm w}_0$ and for ${\rm w}_a$.}
  \label{fig:f5}
\end{figure}

\begin{figure}[t]
\resizebox{0.4\textwidth}{!}{%
 \includegraphics{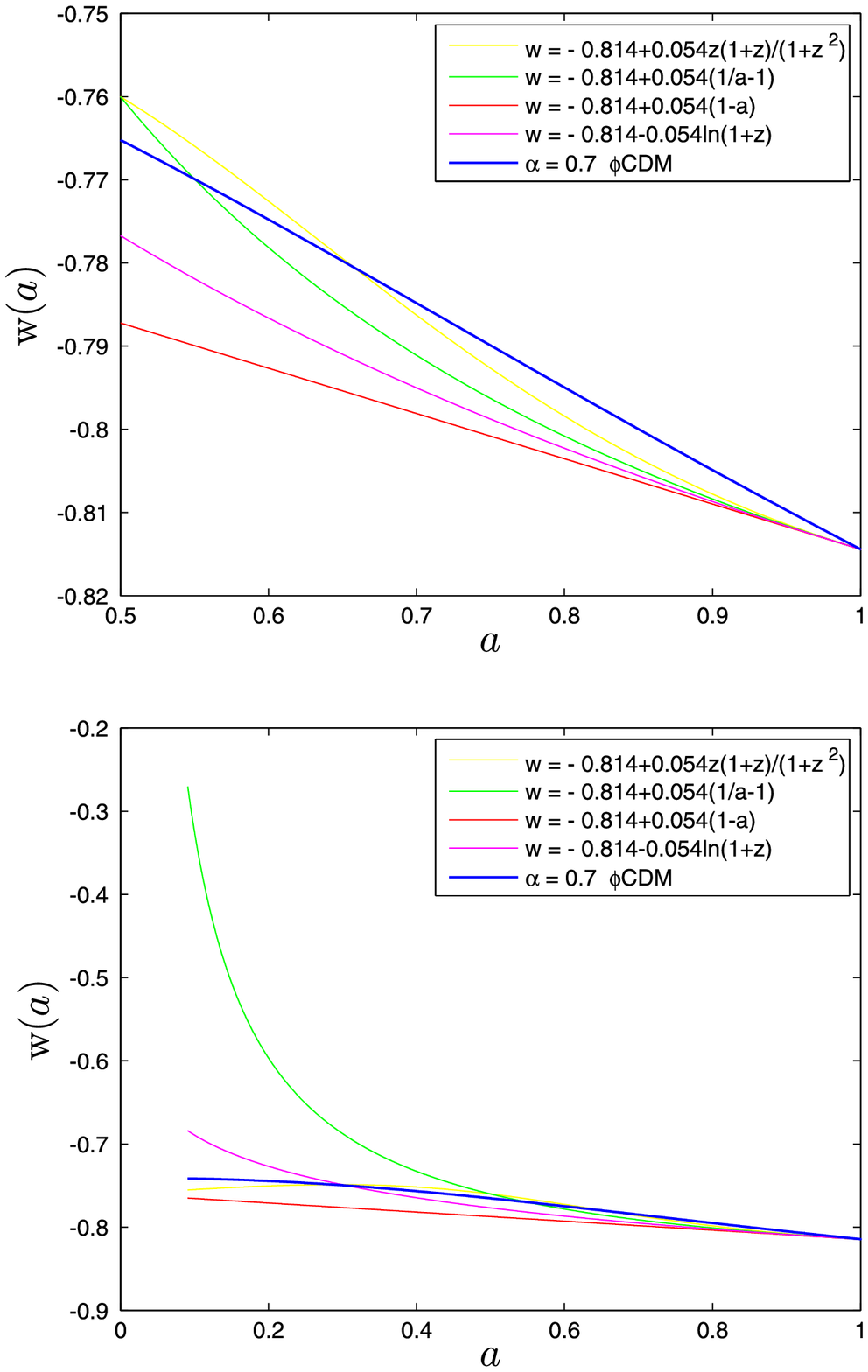}}
 \caption{6a (upper panel) CH, GE, CPL, and BA parameterizations of the ${\rm w}(a)$ and  ${\rm w}(a)$ for the Ratra-Peebles $\phi$CDM model, $\alpha=0.7$ in the range of the redshifts  ${\rm z}\in(0;1)$. 6b (lower panel) BA, CH, GE, CPL parameterizations of the ${\rm w}(a)$  and ${\rm w}(a)$ for the Ratra-Peebles $\phi$CDM model, $\alpha=0.7$ in the range of the redshifts ${\rm z}\in(0;5)$.}
 \label{fig:f6}
\end{figure}

The parametrization of the equation of state parameter ${\rm w}(a)$ is used as a
method to distinguish the different DE models among themselves \citep{scherrer15}. In particular,
this approach can be used to distinguish the $\Lambda$CDM and the $\phi$CDM models at present moment.
We present our results on the Fig.~\ref{fig:f4} - Fig.~\ref{fig:f5}. BA and CH parameterizations fit better the equation of state parameter ${\rm w}(a)$ in  the Ratra-Peebles $\phi$CDM model in the range of the redshifts ${\rm z}\in(0;1)$ then GE and CPL parameterizations.  The  CH, GE, CPL, BA  ${\rm w}(a)$ parameterizations and the function ${\rm w}(a)$  in the Ratra-Peebles $\phi$CDM model, for $\alpha=0.7$ in the range of the redshifts ${\rm z}\in(0;1)$ are represented on  the Fig.~\ref{fig:f6}a. But for the early epochs (large redshifts), %the ${\rm w}(a)$ parameterizations CH and GE are the divergent functions herewith
the CPL and BA ${\rm w}(a)$ parameterizations  approximate better the equation
of state parameter  ${\rm w}(a)$ in the Ratra-Peebles $\phi$CDM model for the
large redshifts then BA and CH parameterizations  (see the Fig.~\ref{fig:f6}b).

We evaluate the Universe expansion rate for the BA, CH, GE and CPL parameterizations  \citep{carroll04}:
\begin{equation}
E^2(a)= \Omega_{\rm m0} a^{-3} + \Omega_{\rm \phi0}e^{-3\int_1^a \frac{{\rm d} a}{a}(1+{\rm w}(a))},
\label{eq:Friedmann_compare}
\end{equation}
 and we compare them with the value of the expansion rate for the Ratra-Peebles
$\phi$CDM model. The expansion rates  for  all aforementioned parameterizations
of ${\rm w}(a)$ fit well the expansion rate for the Ratra-Peebles $\phi$CDM
model for $\alpha=0.7$ in the range of the redshifts ${\rm z}\in(0;0.6)$  (see
Fig.~\ref{fig:f2}b).

\section{Growth factor of matter density perturbations in the Ratra-Peebles $\phi$CDM model}
\label{sec:growth}

In this section we investigate the influence of the scalar field (of the Ratra-Peebles $\phi$CDM
model) the growth of structure
\citep{pavlov13,tad14}. We integrate numerically  the linear perturbation
equation Eq.~(\ref{eq:deltaeq}) with respect to the relative  density contrast
$\delta \equiv \delta\rho_m/ \overline{\rho}$, where $\overline{\rho}$ and
$\delta\rho_m$ - are the mean density and overdensity of the matter component
respectively.  We apply the initial conditions: $\delta(a_{\rm
in})=\delta^{'}(a_{\rm in})=5\cdot10^{-5}$ \citep{pwb10,cfkmr}.
\begin{equation}
\delta^{''}+\Bigl(\frac{3}{a}+\frac{E^{'}}{E}\Bigr)\delta^{'}-\frac{3\Omega_{\rm {m0}}}{2a^{5}E^{2}}\delta=0,
\label{eq:deltaeq}
\end{equation}
\noindent
 Eq. (\ref{eq:deltaeq}) describes completely  the dynamical evolution of matter perturbations, assuming that the perturbed fluid is a perfect one.
We study the evolution of the perturbations through the linear growth factor $D(a)=\frac{\delta(a)}{\delta(a_0)}$, where $\delta(a_0)$ - is a  value of the density contrast today. We  normalize the linear growth factor $D(a)$ to be unity today, i.e. $D(a_{\rm 0})=1$.

\begin{figure}[t]
\resizebox{0.4\textwidth}{!}{%
 \includegraphics{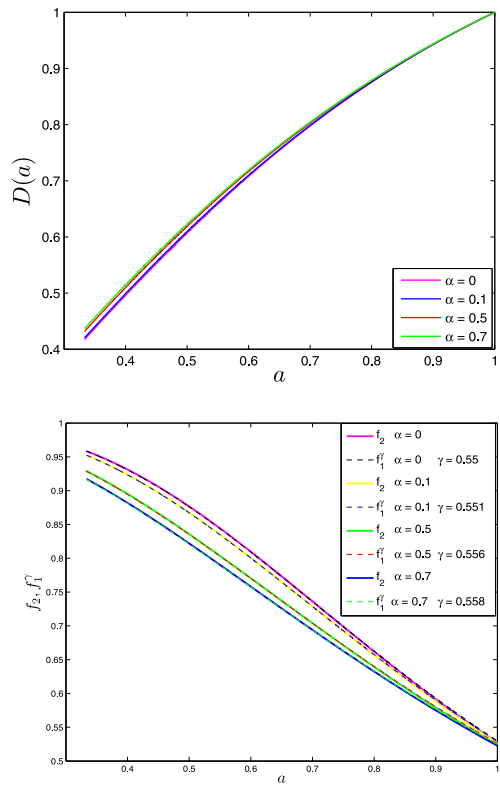}}
\caption{ 7a (upper panel) the linear growth D(a) as a function of the scale factor for different values of the $\alpha$ parameter. 7b (lower panel) the logarithmic growth rate as a function  of the scale factor $f_2$ for different values of the $\alpha$ parameter (solid lines) along with the predictions $f_1^\gamma$ (dashed lines), computed for the corresponding best-fit values of the $\gamma$ parameter.}
\label{fig:f7}
\end{figure}

For the Ratra-Peebles $\phi$CDM model a larger value of the  $\alpha$ parameter
implies a stronger time dependence of the  linear growth factor $D(a)$, see the
Fig.~\ref{fig:f7}a. This is because the growth of matter perturbations occurs
only during  matter dominated epoch \citep{fr08}. The Hubble expansion takes
place faster for larger values of the $\alpha$ parameter (see the
Fig.~\ref{fig:f2}a), while the scalar field energy domination begins earlier,
see the Fig.~\ref{fig:f3}b. As a result, the matter perturbations have less time
to grow. To reach the same amplitude of matter perturbation $\delta(a_0)$
today, the models where the scalar field has a larger value of $\alpha$ will
require larger initial amplitudes.

The growth rate $f_2(a) ={{\rm d}\ln D(a)}/{{\rm d}\ln a}$ strongly depends on
the fractional matter density $f_1(a)=\Omega_{\rm m0}a^{-3}/E^2 $ and this
dependence can be parametrized by a power-law relationship \citep{ws98}
\begin{equation}
 f_2(a) \approx [f_1(a)]^{\gamma},
 \label{eq:f1f2}
\end{equation}
where $\gamma$ is a growth index, the value of which depends on the model parameters.
Assuming GR is correct, the value of $\gamma$-parameter depends on dark energy
parameters as \citep{linder05}.
\begin{equation}
\gamma=0.55+0.05(1+{\rm w}_0+0.5{\rm w}_a),
~~{\rm if}~~ {\rm w}_0 \ge-1.
\label{eq:gamma}
\end{equation}
\noindent
For the $\Lambda$CDM model (with ${\rm w}_a=0$, ${\rm w}_0=-1$), the growth
index $\gamma \approx 0.55.$

We study the applicability of the the power-low approximation
Eqs.~(\ref{eq:f1f2}) for the Ratra-Peebles $\phi$CDM model. This approximation
works well for the Ratra-Peebles $\phi$CDM model (see Fig.~\ref{fig:f7}b). The
value of the growth index $\gamma$ for the $\phi$CDM model depends on the
$\alpha$ parameter. The value of the growth index $\gamma$  for the $\phi$CDM
model increases with increasing value of the $\alpha$ parameter, and it is
slightly higher than one for the $\Lambda$CDM model ($\gamma \approx 0.55$). The
growth rate of matter perturbations occurs slower with the increasing value of
the $\alpha$ parameter  (see Fig.~\ref{fig:f7}b). This results that the Hubble expansion and the growth of matter perturbations are interrelated and oppositely directed processes, and the faster Hubble expansion for the larger $\alpha$ parameter (see the Fig.~\ref{fig:f2}a) leads to the suppression of the growth of matter perturbations.

\begin{figure}[t]
\resizebox{0.4\textwidth}{!}{%
 \includegraphics{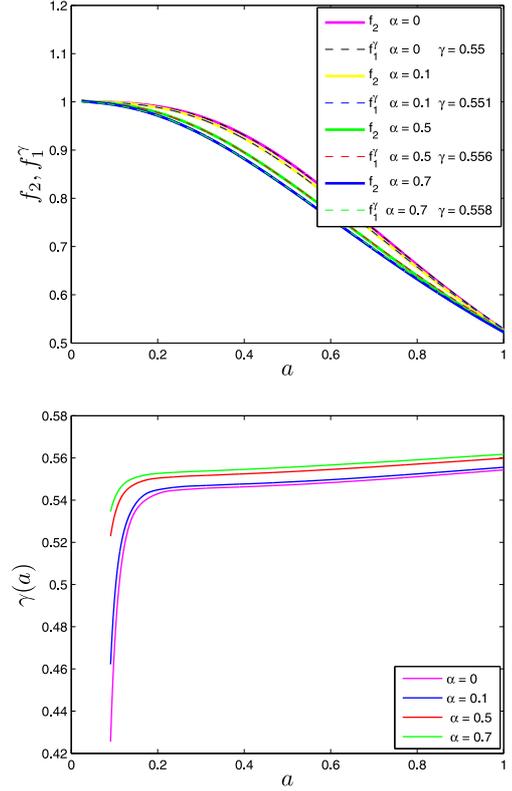}}
 \caption{9a (upper panel) the logarithmic growth rate as a function  of the scale factor $f_2$ for the different values of the $\alpha$ parameter (solid lines) along with the predictions $f_1^\gamma$ (dashed lines), computed for the corresponding best-fit values of the  $\gamma$ parameter in the range of redshifts ${\rm z}\in(0;1)$. 9b (lower panel) the $\gamma(a)$ function for different values of the $\alpha$ parameter in the range of the redshifts ${\rm z}\in(0;10)$.}
 \label{fig:f9}
\end{figure}

We examine also the applicability of the Linder $\gamma$-parametrization on the
large redshifts. This parametrization  occurs in the range of the redshifts
${\rm z}\in(0;5)$  (see Fig.~\ref{fig:f9}a) and for the larger values of the redshifts  the Linder $\gamma$-parametrization is not applicable.

\begin{figure}[t]
\resizebox{0.4\textwidth}{!}{%
 \includegraphics{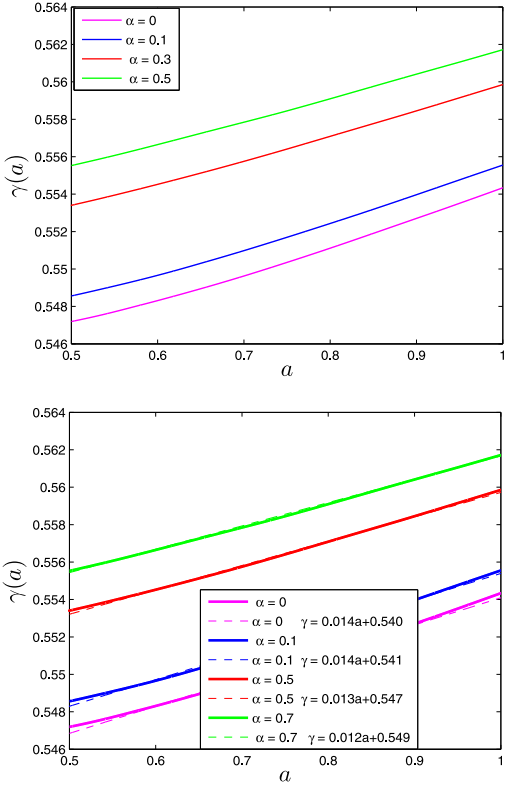}}
 \caption{8a (upper panel)  the $\gamma(a)$ function for the different values of the $\alpha$ parameter in the range of the redshifts ${\rm z}\in(0;1)$.
 8b (lower panel) the $\gamma(a)$ function  for the different values of the $\alpha$ parameter along with predictions, computed from the linear parameterizations.}
 \label{fig:f8}
\end{figure}

We consider also the evolution of the $\gamma(a)$ function with its dependence on the scale factor $a$ \citep{wyf09} for the Ratra-Peebles $\phi$CDM model. From Eqs.~(\ref{eq:f1f2}) we have \citep{linder07,bas14,meh15}:
$$\gamma(a) = \frac{\ln f_2(a)}{\ln f_1(a)}.$$
The $\gamma(a)$ functions in the range of the redshifts ${\rm z}\in(0;1)$ for
different values of the  $\alpha$ parameters are represented on the
Fig.~\ref{fig:f8}a. We approximate the $\gamma(a)$  function  for different
values of the $\alpha$ parameters by the linear functions (see
Fig.~\ref{fig:f8}b).

Next, we investigate the behavior of the $\gamma(a)$ function on the large redshifts, see  Fig.~\ref{fig:f9}b. The linear dependence of the $\gamma(a)$ function  breaks off  in the range of the redshifts from ${\rm z}\approx 3$ for $\alpha=0$ till ${\rm z}\approx 5$ for $\alpha=0.7$. Thus the linear dependence of the $\gamma(a)$ function breaks off earlier for the $\Lambda$CDM model then for the $\phi$CDM models. Comparing the Fig.~\ref{fig:f9}a and the Fig.~\ref{fig:f9}b we see, that the termination of the applicability of the Linder $\gamma$-parametrization for different values of the $\alpha$ parameter  coincides with the moments of the termination of the linear dependence of the $\gamma(a)$ function.

\begin{figure}[t]
\resizebox{0.4\textwidth}{!}{%
 \includegraphics{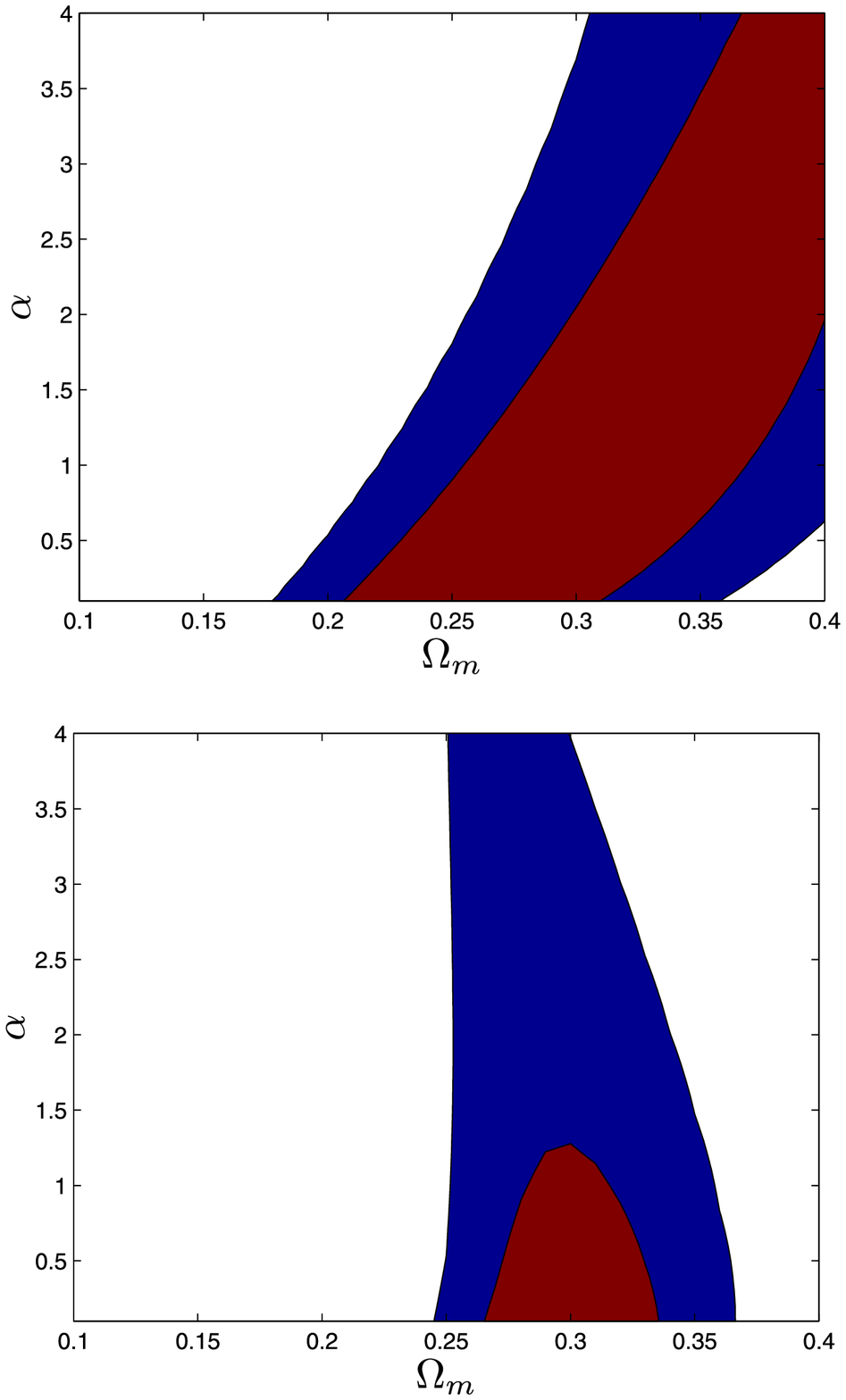}}
 \caption{ 1$\sigma$  and  2$\sigma$  confidence level contours on the parameters
$\Omega_{\rm m}$ and $\alpha$ for the Ratra-Peebles $\phi$CDM model. 10a (upper panel) the
constraints, obtain from the growth rate data \citep{gss12}. 10b (lower panel) the constraints, obtained after adding the BAO
measurements and the cosmic microwave background (CMB) distance prior as in \citep{gswrcl12} for the BAO/CMB distance prior.}
 \label{fig:f10}
\end{figure}

\section{Comparison with observations}
\label{sec:observ}

In this section we carry out the observational constraints on the $\alpha$ and the $\Omega_\mathrm{m}$ parameters using $\chi^2$ analysis, where calculated values of the growth rates are compared with the observational ones, obtained from the redshift space distortion surveys.
For this purpose we use a compilation of the growth rate measurements from
\citep{gss12}.

The 1 and 2$\sigma$ confidence
contours resulting from this likelihood are presented on the Fig.~\ref{fig:f10}a. The likelihood contours in the $\alpha$ - $\Omega_{\rm m}$ plane
obtained from the growth rate data alone are highly degenerate. If we fix
$\alpha=0$ we get $\Omega_\mathrm{m} = 0.278 \pm 0.03$ which is within
1$\sigma$ of the best-fit value obtained by Planck collaboration
\citep{ade13}.
To break the degeneracy between the $\Omega_\mathrm{m}$ and the $\alpha$ parameters we carry out the $\chi^2$ BAO/CMB analysis \citep{gswrcl12}. Where we construct the ratio of the angular distance $d_\mathrm{A}$  and the distance scale $D_\mathrm{V}$:
 $$\eta(z) \equiv d_\mathrm{A}(z_\mathrm{bao})/D_\mathrm{V}(z_\mathrm{bao}),$$
 Assuming Gaussianity of the errorbars
we compute
\begin{equation}
\chi^2_\mathrm{bao} = \bm{X}^\mathrm{T}\bm{C}^{-1}\bm{X},
\end{equation}
\noindent
and a likelihood function
\begin{equation}
\mathcal{L}^\mathrm{bao}(\alpha,\Omega_\mathrm{m},H_0) \propto \mathrm{exp}(-\chi^2_\mathrm{bao}/2),
\end{equation}
\noindent
where $\bm{X} = \eta_\mathrm{th} - \eta_\mathrm{m}$ and $\bm{C}$ is the
covariance matrix of the measurements.  To marginalize over parameter $H_0$ in
$\mathcal{L}^\mathrm{bao}$ we take a Gaussian prior of $H_0 = 74.3 \pm 2.1$
from \citep{freedman12}. We assume that $\mathcal{L}^\mathrm{f}$ and
$\mathcal{L}^\mathrm{bao}$ are independent and the combined likelihood is
simply a product of the two. The results are presented on the Fig.\ref{fig:f10}b.

As a result of the BAO/CMB  analysis we've obtained the  restrictions  on the parameters $\Omega_{\rm m}$ and $\alpha$. $\Omega_{\rm m}$ is now constrained to be within $0.26 < \Omega_{\rm m} < 0.34$ at 1$\sigma$ confidence level. For $\alpha$ parameter we
get $0\leq\alpha\leq1.3$ at 1$\sigma$ confidence level.

\section{Discussion and Conclusions}
\label{sec:conclusion}
 Analyzing the obtained results we can conclude that the Ratra-Peebles $\phi$CDM model differs from the  $\Lambda$CDM model in number of ways. These distinctive features  are generic and do not depend on the specific values of model parameters.
In the Ratra-Peebles $\phi$CDM model the expansion rate of the Universe is always faster than for the $\Lambda$CDM model. The DE dominated epoch sets in earlier than for the $\Lambda$CDM model. The scalar field model predicts a slower growth rate than the $\Lambda$CDM model.

Below we summarize our results:

  We've investigated the parametrization of the equation of state parameter ${\rm w}(a)$ for the Ratra-Peebles $\phi$CDM model by the  Coorkay-Huterer,
  Gerke-Estathiou, Chavalier-Polarsry-Linder, and Barboza-Alcaniz models. We've found that the Barboza-Alcaniz and Coorkay-Huterer parameterizations fit well the equation of state parameter ${\rm w}(a)$ in the Ratra-Peebles $\phi$CDM model in the range of the redshifts ${\rm z}\in(0;1)$. While the Gerke-Estathiou and Chavalier-Polarsry-Linder parameterizations fit well the equation of state parameter ${\rm w}(a)$ for the Ratra-Peebles $\phi$CDM model on the large redshifts.

  The expansion rates calculated for  all  aforementioned parameterizations of ${\rm w}(a)$ fit well the expansion rates for the Ratra-Peebles $\phi$CDM model for all values of the $\alpha$ parameter in the range of the redshifts ${\rm z}\in(0;0.6)$.

   We've explored the Linder $\gamma$-parametrization for the Ratra-Peebles $\phi$CDM and the $\Lambda$CDM models. The Linder $\gamma$-parametrization works well for both models and it is applicable in the range of the redshifts ${\rm z}\in(0;5)$.

   We've found that the effective $\gamma(a)$ function can be  approximated as a linear function in the  range of the redshifts  ${\rm z}\in(0;5)$, which coincides with the range of the redshifts  applicability of the  Linder $\gamma$-parametrization.

We've explored the observable predictions of the scalar field model. We've used a
compilation of the BAO, the growth rate and the distance prior from the CMB to constrain the
model parameters of the scalar field model. When only the growth rate data is
applied, there is a strong degeneracy between the $\Omega_m$ and the $\alpha$ parameters.

 Adding the BAO data and the distance prior from the CMB brake the degeneracy resulting
in $\Omega_m = 0.3$ and $\alpha \leq 1.3$ with the best fit of $\alpha = 0.$

{\it Acknowledgements}
We appreciate useful discussions with Leonardo Campanelli, Bidzina Chargeishvili, Gennady Chitov, Vasil Kukhianidze, Tengiz Mdzinarishvili, Anatoly Pavlov, Bharat Ratra,  Robert Scherrer, and Alexander Tevzadze.  We
acknowledge partial support from Shota Rustaveli Georgian National Science Foundation grant FR/264/6-350/14, the Swiss NSF grant SCOPES IZ7370-152581, the
CMU Berkman foundation, the NSF grants AST-1109180, and the program "The origin, structure and evolution of objects of the Universe" of the Presidium of Russian Academy of Sciences (P-41).

%\vskip10pt
%\bibliography{ourbib}

\end{document}